\documentclass[11pt]{article}
\usepackage{graphicx}
\usepackage{moriond,epsfig}
\begin{document}
\vspace*{4cm}
\title{$CP$-violation and $\pi\pi$-interaction in the radiative decays of
$K_L$ and $K_S$.}
\author{S. S. Bulanov}
\address{ Institute of Theoretical and Experimental Physics,\\
 B. Cheremyshkinskaya ul., 25, 117259 Moscow, Russia.\\
 e-mail: bulanov@heron.itep.ru}

\date{}
\maketitle

\abstracts{ The phases of terms of
$K_{L,S}\rightarrow\pi^{+}\pi^{-}\gamma$ decay amplitude that
arise from the $\pi\pi$ interaction are obtained by using a simple
realistic model of $\pi\pi$ interaction via virtual $\rho$-meson,
instead of the ChPT. It is shown that the standard ChPT approach
cannot reproduce the contribution of the $\rho$-meson to the
$\pi\pi$ interaction. The interference between the terms of
amplitude with different $CP$-parity appears only when the photon
is polarized (linearly or circularly). Instead of measuring the
linear polarization, the angular correlation between the
$\pi^{+}\pi^{-}$ and $e^{+}e^{-}$ planes in
$K_{S,L}\rightarrow\pi^{+}\pi^{-}e^{+}e^{-}$ decay can be
studied.}

\section{Introduction}

The theoretical and experimental study of the $CP$-violation in
the radiative decays of the $K_L$ and $K_S$ has a long history. In
view of future precise measurements of these decays we have
recalculated the above effects. Generally the results that will be
presented in this talk are in agreement with the previous ones. A
few discrepancies are caused by more realistic evaluation of
phases caused by the $\pi\pi$-interaction in the $K_L$ and $K_S$
decays. More accurate calculation of these phases is the
prerequisite for extracting the precise values of the
$CP$-violating parameters in the $K$-meson decays.

The pattern of the CP-violation in the $K_{L,S}\rightarrow \pi
^{+}\pi ^{-}\gamma$ decays was theoretically predicted in the
1960s \cite{bib Chew}, \cite{bib 2a}, \cite{SW}, \cite{bib
dolgov}. In the 1990s these decays were thoroughly studied using
ChPT \cite{AmbrI}, \cite{bib valencia}. The $K_L\rightarrow \pi
^{+}\pi ^{-}\gamma$ decay attracted special attention, because the 
contributions of the $K_L$ decay amplitude terms with opposite $CP$-parity are 
of comparable magnitude and this makes the $CP$-violation to be distinctively
seen experimentally \cite{bib carrol}, \cite{bib ramberg} with the result 
$Br(CP=+1)=(1.49\pm 0.08)\times 10^{-5}$, $Br(CP=-1)=(3.19\pm 0.16)\times 
10^{-5}$. Contrary to this in the case of the $K_S$ decay the $CP$-violation 
is difficult to detect due to the fact that the internal
bremsstrahlung contribution shades the contribution of the direct
emission \cite{bib ramberg}.

In this talk I present the results of the calculations of the
phases of the amplitude terms, connected with the
$\pi\pi$-interaction, using instead of ChPT a simple realistic model
of $\pi\pi$-interaction via virtual $\rho$-meson \cite{bib Lee}. I compare my
results on the photon energy dependence of the
interference between the internal bremsstrahlung and the electric
direct emission in the $K_S$ decay \cite{bulanov} with the results of Ref.
\cite{AmbrI} obtained in the framework of ChPT. It
is proved that the "interference branching ratio" differs from
the the one obtained in Ref. \cite{AmbrI} (see Table
1) and that the standard ChPT approach cannot reproduce the contribution of
the $\rho$. In order to show that the inclusion of the
$\rho$ into the analysis of the $K_{S,L}$ decays is
important, I address the problem of $P$-wave $\pi\pi$ scattering. I compare
the experimental data with the results obtained in the framework of
different models: a) the standard ChPT, b) the ChPT with $\rho$ and
c) the simple realistic model.

According to the approach proposed in Ref. \cite{bib Sehgal1} for
$K_L$ decay, I examine the $K_S\rightarrow \pi^{+}\pi^{-}\gamma$
decay probability with the polarized photon, taking into account
various cases of the photon polarization. It is known that an
alternative to measuring the linear photon polarization, the
angular correlation of $\pi^{+}\pi^{-}$ and $e^{+}e^{-}$ planes in
$K_{S,L}\rightarrow\pi^{+}\pi^{-}e^{+}e^{-}$ decay can be studied.
I present the result of the calculations of the $CP$-violating
asymmetry in the case of the $K_L$ decay in order to compare our
results with those of Ref. \cite{Savage1}. I also present the
result for the $CP$-violating asymmetry in the case of the $K_S$
decay.

\section{The amplitude structure}

The amplitudes of
$K_{S,L}(r)\rightarrow\pi^{+}(p)\pi^{-}(q)\gamma(k,e)$ decays are
made up of two terms: the internal bremsstrahlung ($B$) and direct
emission ($D$). In its turn, $D$ is a sum of an electric
term($E_{D}$) and a magnetic term($M_{D}$). In accordance with the
above, the amplitudes of the $K_{S,L}$ decays can be written as
follows
\begin{equation}
A(K_S\rightarrow \pi^{+}\pi^{-}\gamma)=
eAe^{i\delta_{0}^0}T_{B}+e(a e^{\delta_1^1}+b e^{\delta_b})T_E
+ie\eta_{+-}c e^{\delta_1^1} T_M,   \label{eq.2}
\end{equation}
\begin{equation}
A(K_L\rightarrow \pi^{+}\pi^{-}\gamma)=
\eta_{+-}eA e^{i\delta_{0}^0}T_{B}
+e\eta_{+-}(a e^{\delta_1^1}+b e^{\delta_b})T_E
+i~e c e^{\delta_1^1}T_M,   \label{eq.3}
\end{equation}
where $T_{B}=\frac{pe}{pk}-\frac{qe}{qk}$,
$T_{E}=(pe)(qk)-(qe)(pk)$,
$T_{M}=\varepsilon_{\mu\nu\rho\sigma}p_\mu q_\nu k_\rho e_\sigma$;
$\delta_{0}^0$ and $\delta_{1}^1$ are the $S$-wave and $P$-wave pion
scattering phases respectively and $\eta_{+-}$ is the well known
CP-violation parameter in the $K_L\rightarrow\pi^{+}\pi^{-}$ decay.
The factor $A\equiv A(K\rightarrow\pi^{+}\pi^{-})$ is determined by
the Low theorem for bremsstrahlung \cite{bib 2}. As follows from Eqs. 
(\ref{eq.2}, \ref{eq.3}) $E_D$ consists of two terms. The first term ($a 
e^{\delta_1^1}$) describes the loops of heavy particles. The second term ($b 
e^{\delta_b}$) describes the loops of pions. Such subdivision is convenient 
because the pion loops contribution has an absorptive part and hence a phase,
contrary to the contribution of the heavy particle loops.

The phases of the pion loops and bremsstrahlung contributions in
equations (\ref{eq.2}), (\ref{eq.3}) are defined by the strong
interaction of pions via virtual $\rho$-meson:
\begin{equation}
L=-\frac{g}{\sqrt{2}}\varepsilon_{ijk}(\phi^i\partial_{\mu}{\phi^j}
-\partial_{\mu}{\phi^i}\phi^j)B^k_{\mu},
\end{equation}
where $i,j,k$ are isotopic indices, $\phi$ is the pion field,
$B^k_\mu$ -- $\rho$-meson field.

\begin{figure}[h!]

\begin{tabular}{cc}
\epsfxsize7cm\epsffile{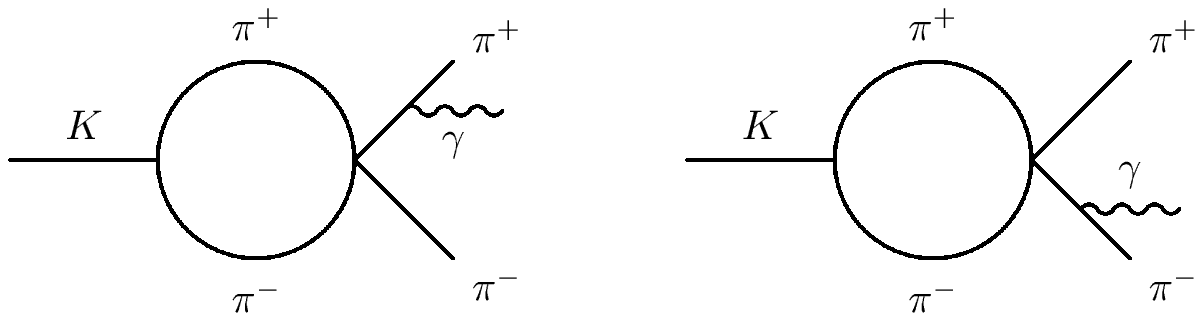} &
\epsfxsize7cm\epsffile{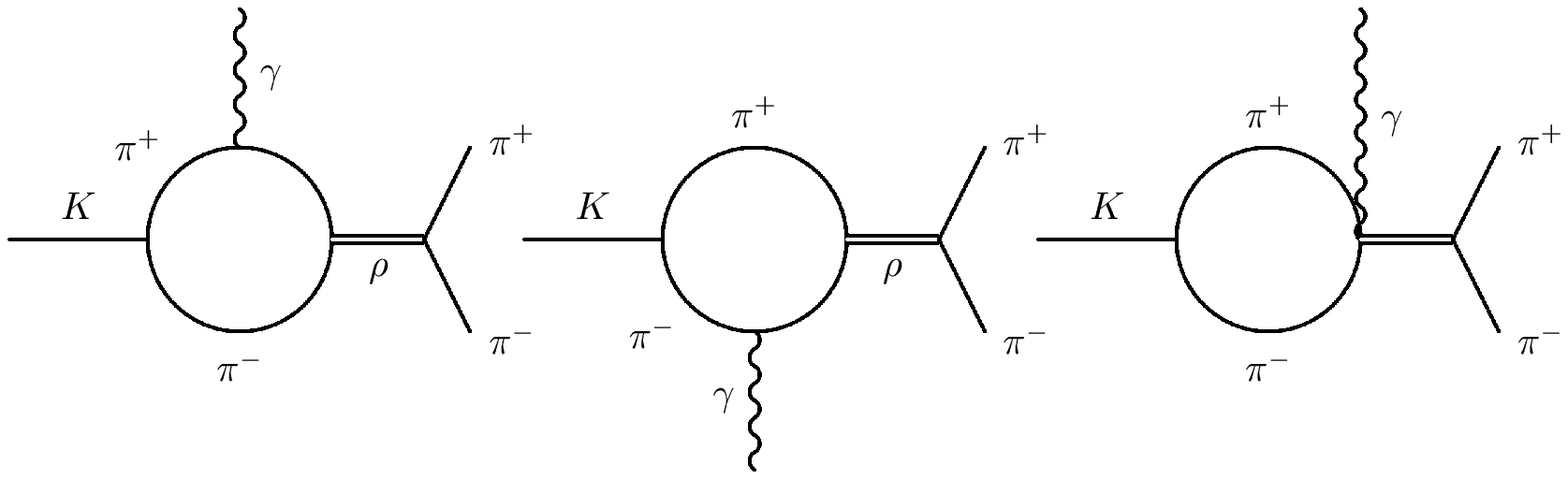}\\
a) & b)
\end{tabular}
\caption{a) The interaction of pions in the case of the internal
bremsstrahlung. b) The emission of the photon from the loops of virtual
particles ($D$).}

\end{figure}

In the case of internal bremsstrahlung contribution to the
$K_{S,L}$ decay probability the interaction of pions can be
described by the diagrams shown in Fig. 1a. Though pions are in
$P$-wave in the final state this group of diagrams results in the
$\delta_0^0$ phase of the amplitude. It is due to the fact that
the interaction of pions occurs not in the final state, but in the
intermediate one.

The emission of the photon from the loops of pions is governed by
another group of diagrams shown in Fig. 1b. Each of these diagrams is
divergent but their sum is finite. The matrix element arising from the
diagrams,
shown in Fig. 1b, takes the form:
\begin{equation}
E_D^{loop}=\frac{eg^2
A}{\pi^2}F(s)D(s)\frac{T_E}{rk}
=ebT_E\label{eq. F},
\end{equation}
with
\begin{eqnarray}
F(s)=\frac{1}{2}+\frac{s}{2rk}\left[\beta K(\beta)-\beta_0
K(\beta_0)\right]-\frac{m_\pi^2}{rk}\left(K^2(\beta)-K^2(\beta_0)\right)
\nonumber \\
+\frac{i\pi}{2rk}
\left(\frac{s}{2}\left(\beta-\beta_0\right)
-2m_\pi^2\left(K(\beta)
-K(\beta_0)\right)\right)
, \label{eq.Fs}
\end{eqnarray}
where $s=(r-k)^2$, $\beta=\sqrt{1-4m_\pi^2/s}$,
$\beta_0=\sqrt{1-4m_\pi^2/m_K^2}$,
$K(\beta)=\mbox{Arth}\left(\frac{1}{\beta}\right)$, $D(s)$ is the
$\rho$ propagator, $g$ is the interaction constant of
$\rho\pi\pi$.

Note that since $F(s)$ is complex, the phase of the loop
contribution is not equal to the pion $P$-wave scattering phase.
Heavy particles in the loop can also contribute to the electric
direct emission amplitude, though they do not produce any
additional phase. The possible intermediate states are $\pi K$,
$K\eta$ and $KK$. However, the $KK$ loop vanishes in the limit
$m_{K^0}=m_{K^+}$.  The contribution of these loops is calculated using the 
assumption that the pseudoscalar mesons ($\pi$, $K$, $\eta$) couple to $\rho$ 
with equal strength, i. e. $g_{\rho K\pi} =g_{\rho
K\eta}=g_{\rho KK}=g$.

\section{Comparison with ChPT}

The electric direct emission manifests itself in the decay
probability mainly through the interference with the internal
bremsstrahlung. Let us compare the results on the "interference
branching ratio" in the $K_S$ decay obtained in Ref. \cite{AmbrI}
with the results obtained in the framework of the simple realistic
model \cite{bulanov}. In the framework of the ChPT the electric
direct emission is a sum of the loop and counterterm
contributions.

\begin{center}
Table 1

\begin{tabular}{|l|l|l|l|l|}
\hline
cut~~in $\omega$& $\omega >20$\ \mbox{Mev} & $\omega >50$\ \mbox{Mev} &
$\omega >100$\
\mbox{Mev}
\\ \hline
$10^6$ \mbox{Interf}~~\cite{bulanov} & -6.3 & -4.8 & -1.7 \\ \hline
$10^6$\mbox{Interf}($k_f=0$)
~~\cite{AmbrI}& -6.2 & -5.0 & -2.0\\ \hline
$10^6$\mbox{Interf}($k_f=0.5$) ~~\cite{AmbrI}& -10.5 & -8.3 &
-3.3\\ \hline $10^6$\mbox{Interf}($k_f=1.0$) ~~\cite{AmbrI}& -14.8
& -11.7 & -4.7\\ \hline $10^6$\mbox{Interf}($k_f=-0.5$)
~~\cite{AmbrI}& -1.9 & -1.6 & -0.6\\ \hline
$10^6$\mbox{Interf}($k_f=-1.0$) ~~\cite{AmbrI}& +2.4 & +1.8 &
+0.7\\ \hline

\end{tabular}
\end{center}

In Table 1 I present interference contributions to the branching ratio of the
$K_S\rightarrow\pi^{+}\pi^{-}\gamma$ decay, for different values of the 
$\omega$ cut, along with the results of Ref. \cite{AmbrI}. In the case of 
$k_f=0$ the "interference branching ratio" obtained here and the one obtained 
in the framework of the ChPT are in agreement for the photon energy cut 
$\omega>20$ MeV. However, for the photon energy cuts $\omega>50$ MeV and 
$\omega>100$ MeV the discrepancy appears. It is due to the fact that the 
photon spectra of these results differ. It can be clearly seen from Fig. 2a, 
where my result (solid curve) along with the results of Ref. \cite{AmbrI} 
(dashed curves, $k_f=-0.5,~0,~+0.5$) for the photon spectra of the
interference contribution are shown. 

If the counterterm contributions are switched on, then the arising
discrepancy is rather large for given in Ref. \cite{AmbrI} values
of the counterterm contributions. It should be stressed that the
counterterm contributions don't depend on the photon energy and
exhibit the behavior different from our results (Fig. 2a). The
discrepancy mentioned above is due to the fact that in Ref.
\cite{AmbrI} the $\rho$ contribution shows up only in low-energy
constants, while the resonance contribution was not considered.
The phase of the electric direct emission amplitude was taken to
be $\delta_1^1(m_K)$, the phase of  $P$-wave $\pi\pi$ scattering
at fixed energy $\sqrt{s}=m_K$. Instead I used the simple
realistic model of $\pi\pi$ scattering via $\rho$-meson taking
into account the energy dependence of the $\delta_1^1(s)$ phase.
As it can be seen from Fig. 2a the standard ChPT approach even
with higher order counterterms taken into account cannot reproduce
the contribution of the $\rho$-meson.

\begin{figure}[h!]
\begin{center}
\begin{tabular}{cc}
\epsfxsize5cm\epsffile{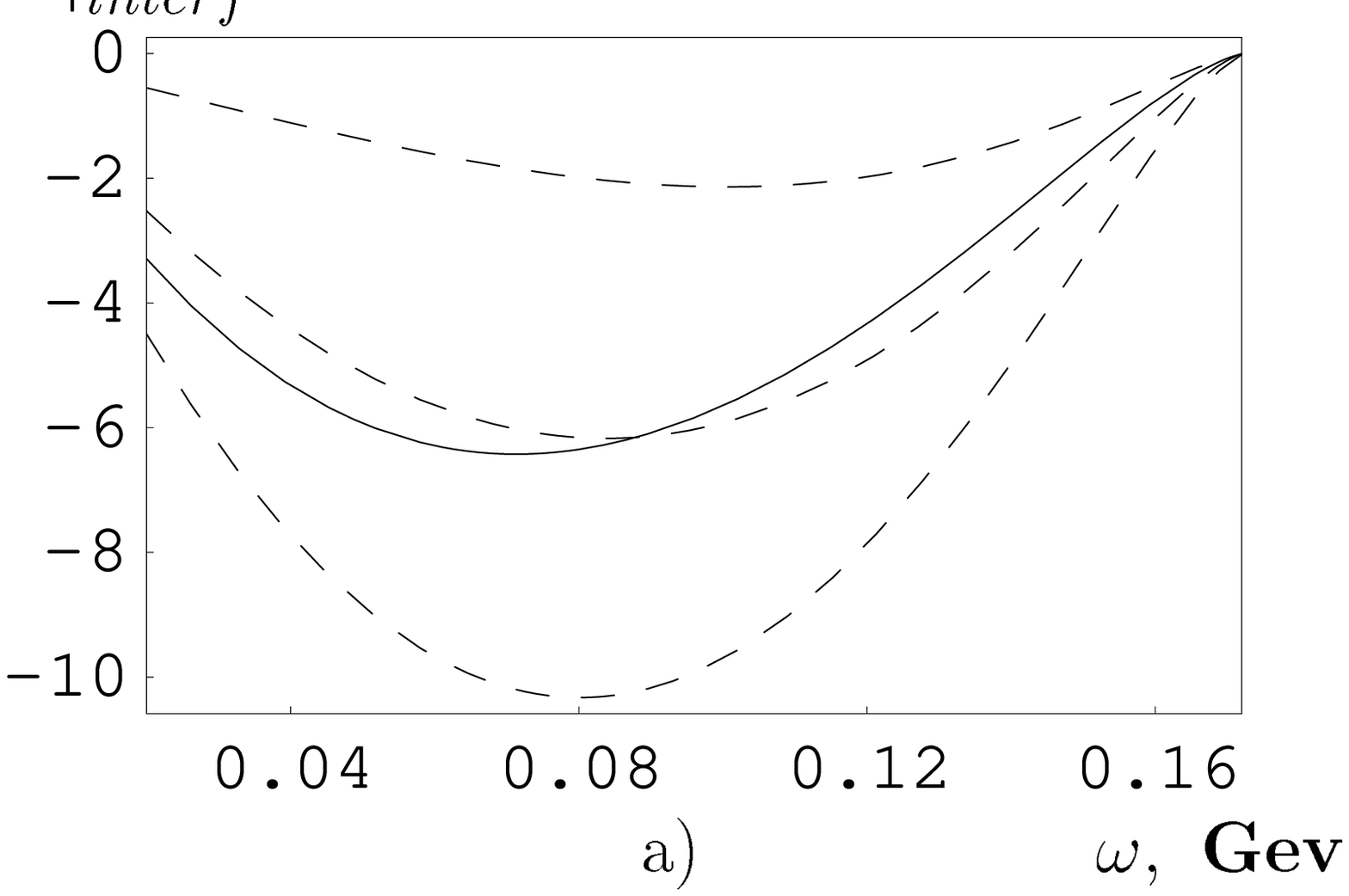} &
\epsfxsize5cm\epsffile{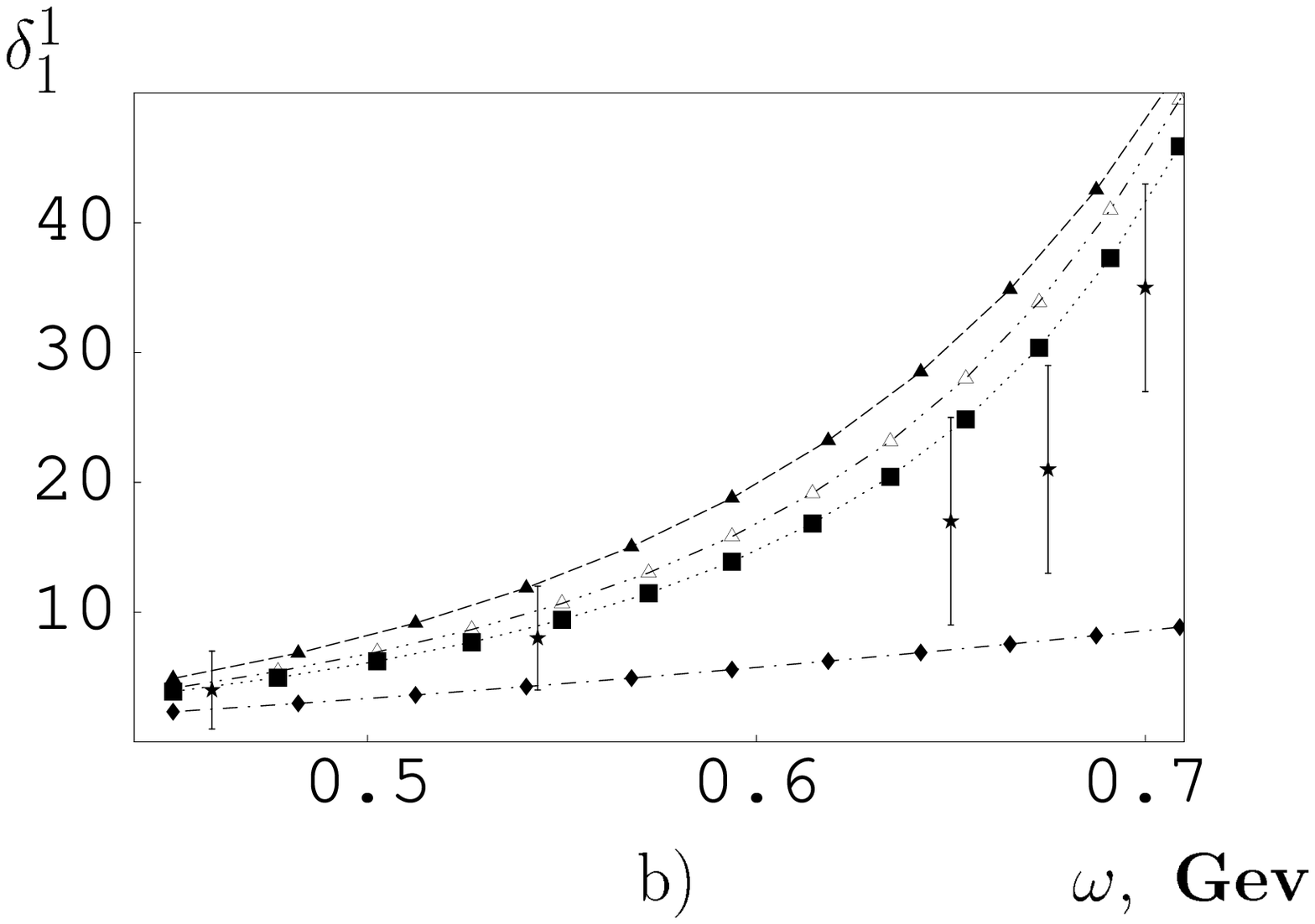}
\end{tabular}
\caption{ a) The photon energy dependence of the interference
contribution. The result of the present paper is represented by
solid curve. The results of ChPT -- by dashed curves. The upper
dashed curve corresponds to $k_f=-0.5$. The lower
dashed curve -- to $k_f=0.5$. The dashed curve in between
-- to $k_f=0$. b) The energy dependence of the $\pi\pi$ scattering
phase $\delta_1^1(s)$ in the $P$-wave, in degrees. The experimental
data are represented by stars. The ChPT without $\rho$
contribution -- by diamonds. The ChPT with the $\rho$
contribution taken into account -- by filled triangles. The ChPT with $\rho$
in two-loop approximation -- by empty triangles. My results -- by boxes.}
\end{center}
\end{figure}

Actually, in the case of $\pi\pi$ scattering in the framework of
the ChPT the $\rho$-meson showed up in two ways: as low-energy
constants and as a direct resonance. However, in papers on the
$K\rightarrow\pi\pi\gamma$ decays only low-energy constants are
accounted for. Such approach doesn't take into account the energy
dependence of the $\delta^1_1(s)$ phase, i. e. the behavior of the
$\rho$ propagator, because the contribution of low-energy
constants doesn't depend on the photon energy. Thus, some
dynamical features are missing in the ChPT approach for the
$K\rightarrow\pi\pi\gamma$ decays.

In order to show that the inclusion of the $\rho$ into the analysis of the
electric direct emission in the $K\rightarrow\pi\pi\gamma$ decays is
important, I address the the problem $P$-wave of $\pi\pi$ scattering. In Fig.
2b I present the behaviour of the $P$-wave $\pi\pi$-scattering phase
$\delta_1^1$ calculated in the framework of different models: a) the standard
ChPT \cite{bib gasser}, b) the ChPT with $\rho$ \cite{bijnens} and
c) the simple realistic model. As it is seen from Fig. 2b the result of the 
standard ChPT is in strict disagreement with the experimental data. It is due 
to the fact that the $\rho$ contribution was not taken into account. The simple
realistic model and the ChPT with $\rho$ are in accordance with the
experimental data. Thus, the inclusion of the $\rho$ in the analysis
of the $K\rightarrow\pi\pi\gamma$ decays is important.

\section{Analysis in terms of Stokes parameters}

It is worth mentioning that there is no interference between the
amplitude terms with opposite $CP$-parity, if photon polarization is
not observed. However, the interference is nonzero when the
polarization is measured. Therefore, any $CP$-violation involving interference
of electric and magnetic amplitudes is encoded in the polarization state of the
photon.

In order to study this interference we write the $K_{L,S}$
decay amplitude more generally as
\begin{equation}
A(K_{S,L}\rightarrow \pi^{+}\pi^{-}\gamma)=ET_E+MT_M,
\label{eq.1}
\end{equation}
where for the $K_S$ decay $E$ and $M$ have the form $
E=eb+eAe^{i\delta_0^0}/(pk~qk)~~\mbox{and}~~
M=ie\eta_{+-}c$, and in the case of the $K_L$ decay we have
$E=\eta_{+-}\left[eb+eAe^{i\delta_0^0}/(pk~qk)\right]~~\mbox{and}~~
M=iec$.

The photon polarization can be defined in the terms of the density matrix
\begin{equation}
\rho=\left(
\begin{array}{c}
|E|^2~~~~E^{*}M\\
EM^{*}~~~~|M|^2
\end{array}
\right)=\frac{1}{2}(|E|^2+|M|^2)\left[1+\bf{S\tau}\right],
\end{equation}
where $\bf{\tau}=(\tau_1,\tau_2,\tau_3)$ are Pauli matrices,
$\bf{S}$ is the Stokes vector of the photon with components
\begin{equation}
S_1=\frac{2Re(E^{*}M)}{(|E|^2+|M^2|)}, ~~
S_2=\frac{2Im(E^{*}M)}{(|E|^2+|M^2|)}, ~~
S_3=\frac{(|E|^2-|M|^2)}{(|E|^2+|M^2|)}.\label{eq.Stokes}
\end{equation}
In order to obtain a quantitative estimate of the $CP$-violation
effects the photon energy dependence of the Stokes vector
components can be studied. In Fig. 3a and 3b  the photon energy
dependence of the $S_1$ (coefficient of an interference term in
case of linear polarization) and $S_2$ (net circular polarization)
in the $K_{L,S}$ decay respectively are shown. In figure 3c I
present the $S_3$  photon energy dependence in $K_{L,S}$ decays to
obtain the estimate of the relative strength of the $CP$-violation
effects in the decays under consideration. The obtained results on
the $K_L$ decay coincide with the results of Ref. \cite{bib
Sehgal1}. Taking into account the physical meaning of the $S_3$ I
conclude that the $CP$-violation effects in $K_S$ decay are
substantially smaller. The reason is that the bremsstrahlung
contribution shades the magnetic direct emission contribution even
for the high photon energies.

\begin{figure}[h!]
\begin{center}
\begin{tabular}{ccc}
\epsfxsize4cm\epsffile{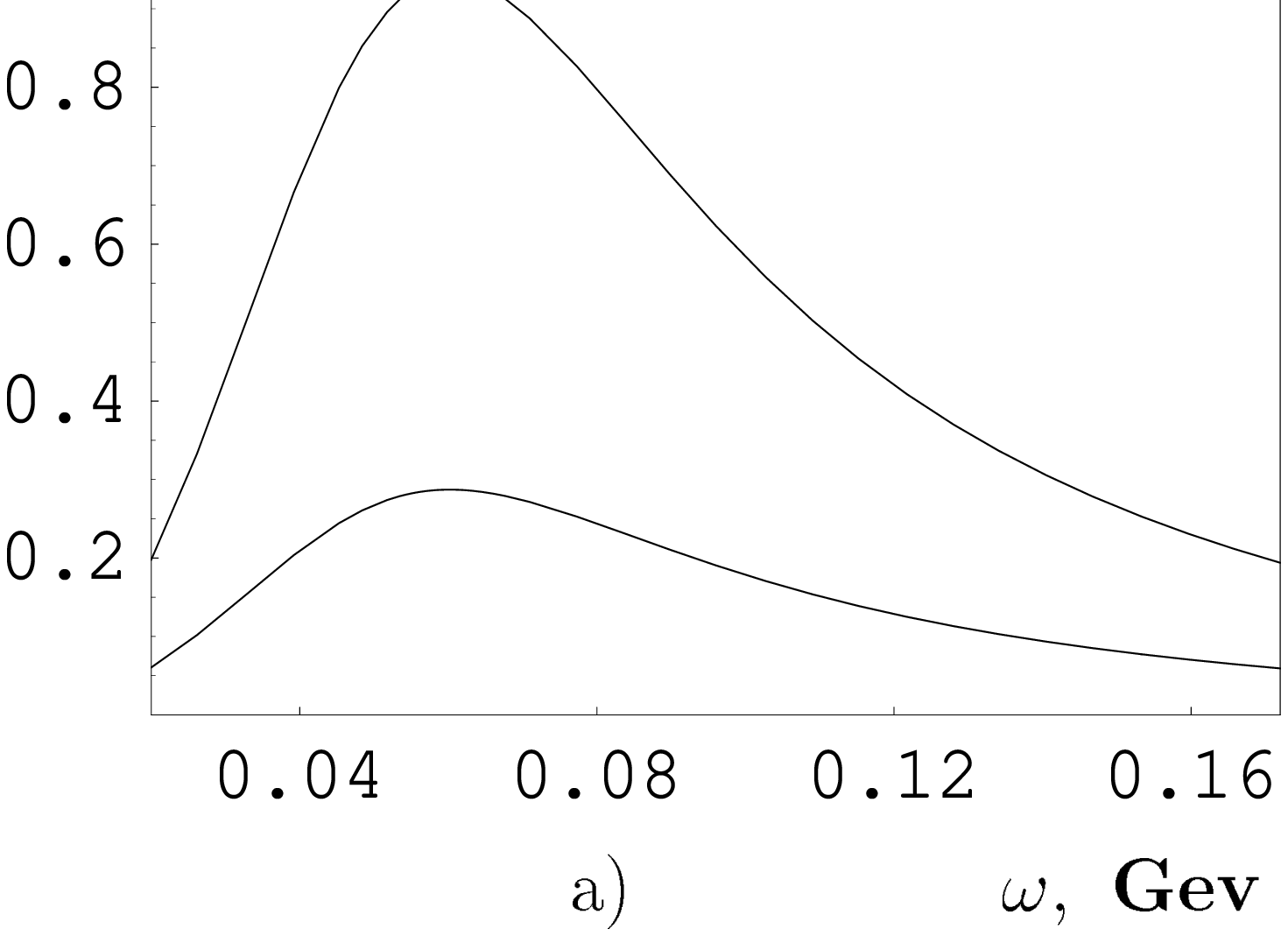} &
\epsfxsize4cm\epsffile{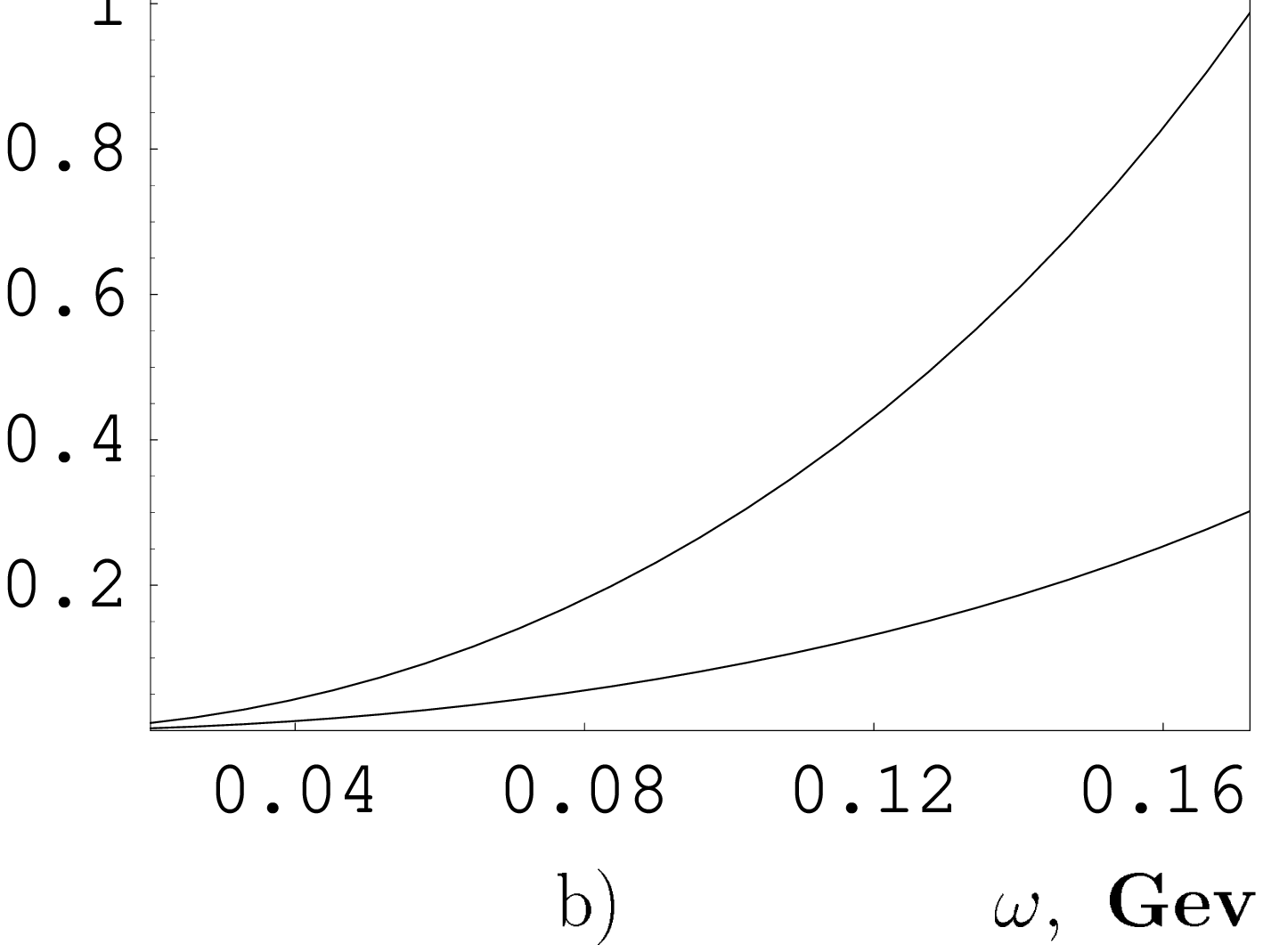} &
\epsfxsize4cm\epsffile{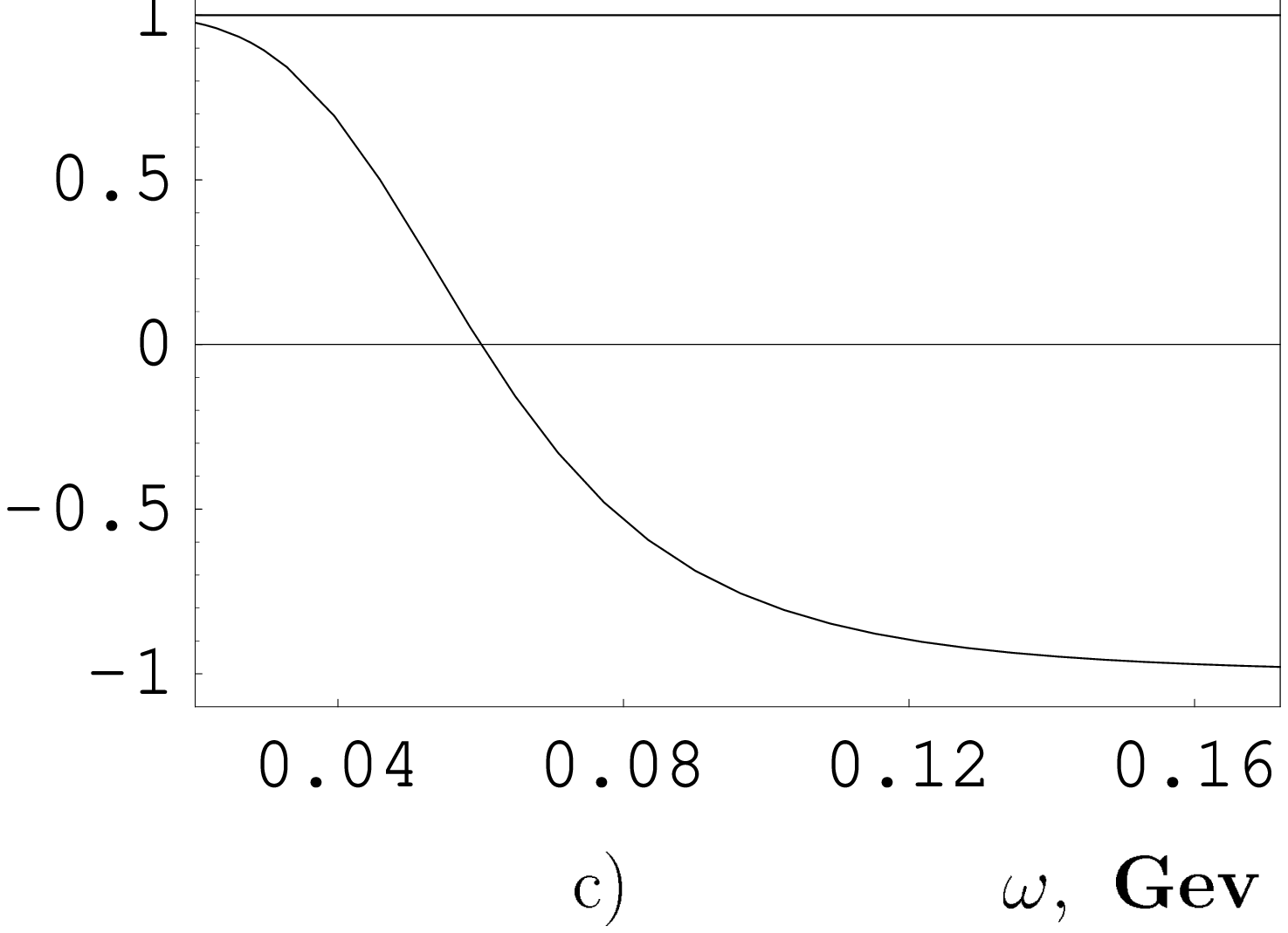}
\end{tabular}
\caption{a) Stokes parameters $S_1$ (upper curve) and $S_2$ (lower
curve) for the $K_L$ decay. b) Stokes parameters $S_1$ (upper curve)
and $S_2$ (lower curve) for the $K_S$ decay. c) Stokes parameter
$S_3$ for the $K_S$ decay (straight line $S_3=1$) and for the $K_L$
decay (lower curve).}
\end{center}
\end{figure}

It was suggested in Refs. \cite{bib sehgal}, \cite{heiliger} to
use in place of $\bf{e}$ ($e_\mu=(0,\bf{e})$), orthogonal to $\bf{k}$ 
($k_\mu=(\omega,\bf{k})$), the vector
$\bf{n}_l$ normal to the $e^{+}e^{-}$ plane in the decay
$K_L\rightarrow \pi^{+}\pi^{-}e^{+}e^{-}$. This can be achieved by
replacing $e_\mu$ in the radiative amplitude (\ref{eq.2},
\ref{eq.3}) by $e/k^2\overline{u}(k_{-})\gamma_\mu v(k_{+})$. This
motivates the study of the distribution $d\Gamma/d\phi$ in the
decays $K_{S,L}\rightarrow \pi^{+}\pi^{-}e^{+}e^{-}$, where $\phi$
is an angle between $\pi^{+}\pi^{-}$ and $e^{+}e^{-}$ planes.

The resulting $CP$-violating asymmetry in the $K_L$ decay is
$|A^L_{\pi\pi,~ee}|=(13.4\pm 0.9)\% $. It coincides within the accuracy of the 
calculation with the theoretical prediction $|A^L_{\pi\pi,~ee}|=14\%$, 
obtained in Refs. \cite{bib Sehgal1}, \cite{bib sehgal},
\cite{heiliger}, \cite{Savage1} and the experimental result $(13.6\pm 
2.5\pm1.5)\%$ \cite{bib ktev}. 

The $CP$-violating asymmetry in the $K_S$ decay is
$|A^S_{\pi\pi,~ee}|=(5.1\pm 0.4)\times 10^{-5}$.
This value of the asymmetry could be expected qualitatively from
the analysis of the $K_S\rightarrow\pi^{+}\pi^{-}\gamma$ decay
amplitude, where the $CP$-violating magnetic direct emission
contribution is substantially small compared to the
$CP$-conserving part of the amplitude.

\section{Conclusions}

In the case of the radiative $K$-meson decays the phases of
amplitude terms using a simple realistic model of pion-pion
interaction \cite{bib Lee} were calculated. Also the pion loop
contribution ($E_D^{loop}$) to the electric direct emission
amplitude was calculated.

I compared my results on the interference contribution to the
$K_S\rightarrow\pi^+\pi^-\gamma$ decay probability with those of
Ref. \cite{AmbrI} and found that the "interference
branching ratio" differs from the the one obtained in Ref.
\cite{AmbrI} (see Table 1). This discrepancy arises
from different models of $\pi\pi$ interaction and the fact that I
took into account the energy dependence of the phases, while in the
standard ChPT approach the phases were taken at the fixed energy
$\sqrt{s}=m_K$. Thus it is clear that the standard ChPT approach for
the $K\rightarrow\pi\pi\gamma$ decays even with higher order
counterterms taken into account cannot reproduce the contribution of
the $\rho$ meson.

In order to show that the inclusion of the $\rho$ meson into the
analysis of the electric direct emission in the
$K\rightarrow\pi\pi\gamma$ decays is important, I addressed the
problem of the $\pi\pi$ scattering in $P$-wave. I compared with the
experimental data the predictions for the phase of the $\pi\pi$
scattering in the $P$-wave obtained in the framework of different
models. As seen from Fig. 2b the simple realistic model and the ChPT
with $\rho$ are in accordance with the experimental data. The ChPT
without $\rho$ shows strong disagreement with data. Thus the inclusion of the
$\rho$ meson, done in the framework of the simple realistic model, in the
$K\rightarrow\pi\pi\gamma$ decays is important.

Regarding the dependence of the $K_S$ decay probability on
photon polarization I found that the effects of $CP$-violation are small
compared to that of the $K_L$ decay, which could be qualitatively expected
from the analysis of the $K_S$ decay. The reason is that the bremsstrahlung
contribution shades the magnetic direct emission contribution even for the
high photon energies.

I also studied the $K_{S,L}\rightarrow\pi^{+}\pi^{-} e^{+}e^{-}$
decays. The central values of the asymmetries presented in this talk, obtained 
in Refs. \cite{bib Sehgal1}, \cite{bib sehgal}, \cite{heiliger}, 
\cite{Savage1} and measured experimentally \cite{bib ktev} coincide within the 
accuracy of the calculation. I found that the $CP$-violating asymmetry in the 
case of the $K_S$ decay is substantially smaller than in the $K_L$ case, as it 
could be expected from the analysis of the 
$K_S\rightarrow\pi^{+}\pi^{-}\gamma$ decay.

The author appreciates L. B. Okun's scientific supervision and
formulation of this problem. The author would also like to thank R.
B. Nevzorov for fruitful discussions and G. D'Ambrosio, M. I.
Vysotsky, E. P. Shabalin for valuable remarks and the Organizing Committee for
the warm and stimulating atmosphere of the Moriond Conference.

\end{document}